# Thermal Spin-Orbit Torque with Dresselhaus Spin-Orbit Coupling

Chun-Yi Xue, Ya -Ru Wang, Zheng-Chuan Wang*

School of Physical Sciences,

University of Chinese Academy of Sciences, Beijing 100049, China.

*wangzc@ucas.ac.cn

**Abstract**

Based on the spinor Boltzmann equation, we obtain a temperature dependent thermal spin-orbit torque in terms of the local equilibrium distribution function in a two-dimensional ferromagnet with Dresselhaus spin-orbit coupling. We also derive the continuity equation of spin accumulation and spin current —the spin diffusion equation in Dresselhaus ferromagnet, which contains the thermal spin-orbit torque under local equilibrium assumption. This temperature dependent thermal spin-orbit torque originates from the temperature gradient applied to the system, it is also sensitive to temperature due to the local equilibrium distribution function therein. In the spin diffusion equation, we can single out the usual spin-orbit torque as well as the spin transfer torque, which is conceded to our previous results. Finally, we illustrate them by an example of spin-polarized transport through a ferromagnet with Dresselhaus spin-orbit coupling driven by temperature gradient, those torques including thermal spin-orbit torque are demonstrated numerically.

## I. Introduction

As a new branch of spintronics, spin-orbitronics has been explored whatever in theories or experiments [1-2], because it can provide an efficient way to manipulate the local magnetic moment in the devices of spintronics via spin-orbit torques [3-5]. Since the discovery of giant magnetoresistance [6] and spin transfer torque [7], nowadays magnetoresistance random access memory (MRAM) driven by spin-polarized current has been designed and realized industrially [8-10]. The first generation MRAM is toggle MRAM [11], which consist of a transistor and a magnetic tunnel junction (MTJ), but this structure brings obvious disadvantages to the toggle MRAM due to the big magnetic field[12]. The second generation MRAM are spin transfer torque MRAM (STT-MRAM) [13] and perpendicular spin transfer torque MRAM (pSTT-MRAM) [14], in which magnetization reversal in STT-MRAM and pSTT-MRAM rely on the spin-polarized electrical current rather than big magnetic field, so they have faster writing speed. However, STT-MRAM depends on thermal activation to start switching [15], so it has an initial latency which restricts its maximum cache speed. Thus, one proposed the third generation MRAM to solve this problem, which is driven by spin orbit torque (SOT). SOT can be employed to switch the magnetization in a system with a broken inversion symmetry. Since SOT need a lower critical current, it has better thermal stability [16]. Recently, Thermal SOT has been observed in experiments [17-18]. Till now, SOT-MRAM is a good candidate of magnetic storage device for better performance.

In 2007, Hatami proposed STT which can be driven by thermal spin current [19]. Similarly, SOT can also be driven by thermal spin current in the process of spin-polarized electron transport, which is called thermal SOT (TSOT). TSOT was firstly proposed by Freimuth in 2014 in terms of Berry phase [20-21] which is expressed by the quantum states of electrons, while the states usually should be calculated by the Density Function Theory (DFT), it is somewhat cumbersome. Thus, in this manuscript we will give another expression of TSOT by use of distribution function in the spinor Boltzmann equation (SBE).

The SBE was firstly proposed by Sheng et al in 1996 at steady state [22], they derived this equation from Kadanoff nonequilibrium Green function (NEGF) formalism based on the gradient approximation [23]. Then SBE was accomplished by Levy et al in 2004, which could illustrate the time dependent process of spin transport effectively [24]. SBE was also extended to the case beyond gradient approximation in 2013 [25]. In 2019, Wang et al. successfully included the Rashba and Dresslhaus spin-orbit coupling into the SBE [26], which is helpful for us to investigate the SOT from the spin diffusion equation – the continuity equation for the spin accumulation and spin current. In this manuscript, our purpose is to investigate TSOT in a 2-dimensional ferromagnet by means of SBE. We will find that an unusual SOT driven by the temperature gradient in the system with Dresselhaus spin orbit coupling, it is just the TSOT we want.

## II. Theoretical Formalism

Consider the spin-polarized electron transport in a two-dimensional ferromagnet with Dresselhaus SOC. It's Hamiltonian can be written as $\widehat{H} = -\frac{\hbar^2}{2m}\vec{\nabla}^2 \widehat{\sigma}_0 + J\vec{M} \cdot \vec{\widehat{\sigma}} + \widehat{H}_D$, where $m$ is electronic effective mass, J is the s-d exchange coupling constant, $\vec{M}$ is the unit vector for the local magnetic moment of ferromagnet, $\widehat{\sigma}_0$ is the unit matrix, $\vec{\widehat{\sigma}}$ is the Pauli matrix vector, $\widehat{H}_D$ describes the Dresselhaus SOC in two dimensional ferromagnet. In 1955, Dresselhaus gave the Hamiltonian of Dresselhaus SOC in three dimensional system with a bulk inversion asymmetry(BIA) [29]. When such a three dimensional system is subjected to strain at the interface or a thin layer, the Hamiltonian will reduce to the following simpler form: $\hat{H}_D = \lambda_D(\hat{\sigma}_x \nabla_x - \hat{\sigma}_y \nabla_y)$ [26], where $\lambda_D$ is the coupling constant of Dresselhaus SOC, $\nabla_x$ and $\nabla_y$ are the x and y components of gradient operator, and the $\hat{\sigma}_x$ and $\hat{\sigma}_y$ are the x and y components of the Pauli operator $\vec{\hat{\sigma}}$. By Kadanoff nonequilibrium Green function (NEGF) formalism, we can obtain the SBE for the spinor distribution of transport electron [24]:

$$\left(\frac{\partial}{\partial t} + \vec{\hat{v}} \cdot \frac{\partial}{\partial \vec{r}} - e\vec{E} \cdot \frac{\partial}{\partial \vec{p}}\right)\hat{f} + \frac{i}{\hbar}[\hat{\varepsilon}, \hat{f}] = -\left(\frac{d\hat{f}}{dt}\right)_{collision} \quad \#(1)$$

where $\hat{f}$ is the spinor distribution function of transport electron, which can be expanded to a 2×2 matrix $\hat{f} = \begin{pmatrix} f_{\uparrow\uparrow} & f_{\uparrow\downarrow} \\ f_{\downarrow\uparrow} & f_{\downarrow\downarrow} \end{pmatrix}$, $\hat{\varepsilon}$ is the spinor energy, it's defined as $\hat{\varepsilon}(p) = \varepsilon(p)\hat{\sigma}_0 + \frac{1}{2}J(p)\vec{M}(x,t)$. In the SBE framework, we can decompose the spinor distribution function into two parts by using the complete basis of matrices $\hat{\sigma}_0$ and the $\vec{\hat{\sigma}}$, $\hat{f}(\vec{R},\vec{p}) = f(\vec{R},\vec{p})\hat{\sigma}_0 + \vec{g}(\vec{R},\vec{p})\vec{\hat{\sigma}}$, where $f(\vec{R},\vec{p})$ is scalar distribution function and $\vec{g}(\vec{R},\vec{p})$ is vector distribution function. Under the local equilibrium assumption, the spinor distribution function can also be written as follow:

$$\hat{f}(\vec{R},\vec{p},t) = \hat{f}^0(\vec{R},\vec{p}) + \left(f^1(\vec{R},\vec{p})\hat{\sigma}_0 + \vec{g}^1(\vec{R},\vec{p}) \cdot \vec{\hat{\sigma}}\right) \quad (2)$$

where $\hat{f}^0(\vec{R},\vec{p})$ is the local equilibrium distribution function. $f^1(\vec{R},\vec{p})$ and $g^1(\vec{R},\vec{p})$ are the nonequilibrium parts of spinor distribution. In the ferromagnet, we take the local equilibrium distribution function as a diagonal matrix $\hat{f}^0(\vec{R},\vec{p}) = \begin{pmatrix} f_{\uparrow\uparrow} & 0 \\ 0 & f_{\downarrow\downarrow} \end{pmatrix}$, here the diagonal components are taken as Fermi distribution functions $f_{\uparrow\uparrow(\downarrow\downarrow)} = \left\{exp\left[\frac{\varepsilon(p)\pm\frac{1}{2}J-\mu(x)}{k_BT(x)}\right]+1\right\}^{-1}$, where $k_B$ is Boltzmann constant, $\mu(x)$ is the chemical potential and $\mu \approx E_F$ (Fermi Energy) at finite temperature. We can expand the local equilibrium distribution based on the complete basis $\{\hat{\sigma}_0, \hat{\sigma}_x, \hat{\sigma}_y, \hat{\sigma}_z\}$, which is $\hat{f}^0(\vec{R},\vec{p}) = \frac{1}{2}(f_{\uparrow\uparrow}+f_{\downarrow\downarrow})\hat{\sigma}_0 + \frac{1}{2}(f_{\uparrow\uparrow}-f_{\downarrow\downarrow})\hat{\sigma}_z$, so the scalar distribution and vector distribution can be rewritten as $f = \frac{1}{2}(f_{\uparrow\uparrow}+f_{\downarrow\downarrow}) + f_1$ and $g = \left\{g_x^1, g_y^1, g_z^1 + \frac{1}{2}(f_{\uparrow\uparrow}-f_{\downarrow\downarrow})\right\}$.

In spintronics, the SBE with Dresselhaus spin-orbit coupling in a two-dimensional magneto-electric system under an external electric field $\vec{E}$ had been given by Chao Yang et al. in 2019 [26]

$$\left(\frac{\partial}{\partial t} + \frac{\vec{p}}{m}\cdot\vec{\nabla} - e\vec{E}\cdot\vec{\nabla}_p\right)f(\vec{R},\vec{p}) + \frac{\lambda_D}{\hbar}\left(\frac{\partial g_x}{\partial x} - \frac{\partial g_y}{\partial y}\right) = -\left(\frac{\partial f}{\partial t}\right)_{collision} \quad \#(3)$$

and

$$\left(\frac{\partial}{\partial t} + \frac{\vec{p}}{m}\cdot\vec{\nabla} - e\vec{E}\cdot\vec{\nabla}_p\right)\vec{g}(\vec{R},\vec{p}) - \frac{J}{\hbar}\vec{M}\times\vec{g}(\vec{R},\vec{p}) + \frac{\lambda_D}{\hbar}\left(\vec{e}_x\frac{\partial f}{\partial x} - \vec{e}_y\frac{\partial f}{\partial y}\right)$$
$$+ \frac{\lambda_D}{\hbar^2}(\vec{e}_x p_x - \vec{e}_y p_y)\times\vec{g}(\vec{R},\vec{p}) = -\left(\frac{\partial \vec{g}}{\partial t}\right)_{collision} \quad \#(4)$$

where $-\left(\frac{\partial f}{\partial t}\right)_{collison}$ and $-\left(\frac{\partial \vec{g}}{\partial t}\right)_{collision}$ represent the collision terms.

Under the relaxation time approximation assumption, we can derive the equations for the scalar distribution function and the vector distribution function based on Eqs. (3) and (4), which

contain the local equilibrium distribution function as:

$$\left(\frac{\partial}{\partial t} + \frac{\vec{p}}{m} \cdot \vec{\nabla} - e\vec{E} \cdot \vec{\nabla}_p\right)\left[\frac{1}{2}(f_{\uparrow\uparrow} + f_{\downarrow\downarrow})\right] + \left(\frac{\partial}{\partial t} + \frac{\vec{p}}{m} \cdot \vec{\nabla} - e\vec{E} \cdot \vec{\nabla}_p\right)f^1$$
$$+ \frac{\lambda_D}{\hbar}\left(\frac{\partial g_x^1}{\partial x} - \frac{\partial g_y^1}{\partial y}\right) = -\frac{f - \langle f \rangle}{\tau} \#(5)$$

$$\left(\frac{\partial}{\partial t} + \frac{\vec{p}}{m} \cdot \vec{\nabla} - e\vec{E} \cdot \vec{\nabla}_p\right)\vec{g} - \vec{M} \times \vec{g} +$$
$$\frac{\lambda_D}{\hbar}\left(\vec{e}_x \frac{\partial\left[\frac{1}{2}(f_{\uparrow\uparrow} + f_{\downarrow\downarrow})\right]}{\partial x} - \vec{e}_y \frac{\partial\left[\frac{1}{2}(f_{\uparrow\uparrow} + f_{\downarrow\downarrow})\right]}{\partial y}\right) +$$
$$\frac{\lambda_D}{\hbar}\left(\vec{e}_x \frac{\partial f^1}{\partial x} - \vec{e}_y \frac{\partial f^1}{\partial y}\right)\frac{\lambda_D}{\hbar} - \frac{\lambda_D}{\hbar^2}(p_x\vec{e}_x - p_y\vec{e}_y) \times \vec{g} = -\frac{\vec{g} - \langle\vec{g}\rangle}{\tau_{sf}}\#(6)$$

where $\tau$ and $\tau_{sf}$ are the relaxation times of electron and spin flip, respectively. Eq. (5) and Eq. (6) are coupled together, we should solve them simultaneously. The physical observables in the spin-polarized transport can be expressed by the solutions of scalar and vector distribution functions. The charge density and charge current are defined as follow,

$$\rho(\vec{R}, t) = e \int f^1(\vec{R}, \vec{p}, t) d\vec{p} \#(7)$$

and

$$\vec{j}(\vec{R}, t) = e \int \vec{v} f^1(\vec{R}, \vec{p}, t) d\vec{p} \#(8)$$

which are the momentum integrals over the scalar distributions. Similarly, the spin accumulation and spin current density are define as

$$\vec{m}(\vec{R},t) = e \int \vec{g}(\vec{R},\vec{p},t)d\vec{p} \quad \#(9)$$

and

$$\vec{j}_m(\vec{R},t) = e \int \vec{v}\vec{g}(\vec{R},\vec{p},t)d\vec{p} \quad \#(10)$$

which are the momentum integrals over the vector distributions. It should be noted that the spin current $\vec{j}_m(\vec{R},t)$ is a tensor. Moreover, we can also define the thermal current density as $\vec{j}_E$

$$\vec{j}_E(\vec{R},t) = e \int \varepsilon \vec{v} f^1(\vec{R},\vec{p},t)d\vec{p} \quad \#(11)$$

where $\varepsilon$ is the scalar energy of electron. So when we get the solutions of scalar and vector distribution functions, we can obtain these above physical observables accordingly.

On the other hand, we can also obtain the continuity equations satisfied by these physical observables. If we integrate the momentum over the Fermi surface on the both sides of Eq. (5) and Eq. (6), we have

$$\frac{\partial \rho}{\partial t} + \vec{\nabla}\cdot\vec{j} = -\int \left(\frac{\partial}{\partial t} + \frac{\vec{p}}{\mu}\cdot\vec{\nabla} - e\vec{E}\cdot\vec{\nabla}_p\right)\left[\frac{1}{2}(f_{\uparrow\uparrow} + f_{\downarrow\downarrow})\right]d\vec{p}$$
$$- \frac{\lambda_D}{\hbar}\left(\frac{\partial m_x}{\partial x} - \frac{\partial m_y}{\partial y}\right) - \frac{\rho - \langle\rho\rangle}{\tau} \quad \#(12)$$

and

$$\frac{\partial \vec{m}}{\partial t} + \vec{\nabla} \cdot \vec{j_m} = \frac{J}{\hbar}\vec{M} \times \vec{m}$$

$$\frac{\lambda_D}{\hbar} \int \left( \vec{e}_x \frac{\partial \left[\frac{1}{2}(f_{\uparrow\uparrow}+f_{\downarrow\downarrow})\right]}{\partial x} - \vec{e}_y \frac{\partial \left[\frac{1}{2}(f_{\uparrow\uparrow}+f_{\downarrow\downarrow})\right]}{\partial y} \right) d\vec{p} - \frac{\lambda_D}{\hbar}\left(\vec{e}_x \frac{\partial \rho}{\partial x} - \vec{e}_y \frac{\partial \rho}{\partial y}\right)$$

$$+ \frac{\lambda_D}{\hbar^2} \int (p_x \vec{e}_x - p_y \vec{e}_y) \times \vec{g}\, d\vec{p} = -\frac{\vec{m} - \langle \vec{m} \rangle}{\tau_{sf}} \#(13)$$

Eq. (12) is just the continuity equation for charge density and charge current, while Eq. (13) is the continuity equation for the spin accumulation and spin current, the latter is also called spin diffusion equation. When the time $t \gg \tau_{sf}$, the equation will arrive at a steady state, then the spin diffusion equation will reduce to

$$\frac{J}{\hbar}\vec{M} \times \vec{m} = \vec{\nabla} \cdot \vec{j_m} + \frac{\lambda_D}{\hbar} \int \left( \vec{e}_x \frac{\partial \left[\frac{1}{2}(f_{\uparrow\uparrow}+f_{\downarrow\downarrow})\right]}{\partial x} - \vec{e}_y \frac{\partial \left[\frac{1}{2}(f_{\uparrow\uparrow}+f_{\downarrow\downarrow})\right]}{\partial y} \right) d\vec{p}$$

$$+ \frac{\lambda_D}{\hbar}\left(\vec{e}_x \frac{\partial \rho}{\partial x} - \vec{e}_y \frac{\partial \rho}{\partial y}\right) - \frac{\lambda_D}{\hbar^2} \int (p_x \vec{e}_x - p_y \vec{e}_y) \times \vec{g}\, d\vec{p} \#(14)$$

From the above steady state equation, we can read out all the torques existing in this spin-polarized transport process. On the left side of this equation, the term $\frac{J}{\hbar}\vec{M} \times \vec{m}$ is just the spin transfer torque given by Levy et al. [27]. On the right hand side side, $\vec{\nabla} \cdot \vec{j_m}$ is the divergence of spin current, which also can make a contribution to the usual STT as shown by Zhang et al [28]. Besides, the term $\frac{\lambda_D}{\hbar}(\vec{e}_x \frac{\partial \rho}{\partial x} - \vec{e}_y \frac{\partial \rho}{\partial y}) - \frac{\lambda_D}{\hbar^2} \int (p_x \vec{e}_x - p_y \vec{e}_y) \times \vec{g}\, d\vec{p}$ corresponds to the usual spin-orbit torque presented by Wang et al. [26], while the temperature dependent term $\frac{\lambda_D}{\hbar} \int \left( \vec{e}_x \frac{\partial \left[\frac{1}{2}(f_{\uparrow\uparrow}+f_{\downarrow\downarrow})\right]}{\partial x} - \right.$

$\vec{e}_y \frac{\partial \left[\frac{1}{2}(f_{\uparrow\uparrow}+f_{\downarrow\downarrow})\right]}{\partial y}\right) d\vec{p}$ is a new term, it is induced by the gradient of local equilibrium distribution function, we refer to this as the thermal SOT. When the gradient of temperature is applied only along x-direction, it can be expressed as:

$$TSOT = -\frac{\lambda_D}{\hbar} \cdot \frac{\partial T}{\partial x} \int \left(\vec{e}_x \left[\frac{1}{2}\left(\frac{exp(\frac{\varepsilon-\mu+\frac{1}{2}J}{k_bT})}{\left[1+exp(\frac{\varepsilon-\mu-\frac{1}{2}J}{k_bT})\right]^2} \cdot \left(\frac{\varepsilon-\mu+\frac{1}{2}J}{k_b}\right) + \frac{exp(\frac{\varepsilon-\mu-\frac{1}{2}J}{k_bT})}{\left[1+exp(\frac{\varepsilon-\mu-\frac{1}{2}J}{k_bT})\right]^2} \cdot \left(\frac{\varepsilon-\mu-\frac{1}{2}J}{k_b}\right)\right)\left(\frac{1}{T^2}\right)\right]\right) d\vec{p}$$

(15)

we can see that it is proportional to the gradient of temperature, which is conceded to the definition of TSOT given by Freimuth et al [20-21], so this term is just the TSOT we search for, it is the central result in this manuscript. In the next, we will evaluate these torques numerically in a ferromagnet with Dresselhaus SOC.

### III. Numerical Results

We consider a two-dimensional ferromagnet with Dresselhaus SOC, where the system is chosen as a rectangular ferromagnet with a geometry of 25×25 nm². The temperature distribution is simply chosen as $T(x) = T_0 + kx$, which is linearly dependent on the position of x component, where $T_0$ is a constant, k is the temperature gradient. From Eq. (15), we can see that the temperature gradient will induce thermal spin-orbit torque.

In order to quantify these torques and currents, we need to solve Eq. (3) combining with (4) simultaneously, because the scalar distribution function and vector distribution function are coupled together in these equations. To simplify calculation, we chose the unit vector of magnetization as a fixed vector $\vec{M} = (0,0,1)$, the equilibrium scalar distribution function is chosen as $\langle f \rangle = \frac{1}{2}(f_{\uparrow\uparrow} + f_{\downarrow\downarrow})$, and the equilibrium vector distribution function is adopted as $\langle \vec{g} \rangle = (\exp\left[iJ(\frac{x}{p_x} + \frac{y}{p_y})\right]M_x, \exp\left[iJ(\frac{x}{p_x} + \frac{y}{p_y})\right]M_y, \exp\left[iJ(\frac{x}{p_x} + \frac{y}{p_y})\right]M_z)$ [22]. The differential equations (3) and (4) are solved by difference method. The physical constant and parameters are listed in Table I, where we adopt the materials parameters of ferromagnet.

Table I. The physical constants and parameters

| Physical constants/parameters | Symbol | Value | Unit |
|---|---|---|---|
| Momentum relaxation time | $\tau$ | $10^{-13}$ | s |
| Spin-flip relaxation time | $\tau_{sf}$ | $10^{-12}$ | s |
| Fermi energy | $E_F$ | 4 | eV |
| Fermi wave vector | $k_F$ | $1.02 \times 10^{10}$ | $m^{-1}$ |
| s-d exchange coupling strength | $J$ | 0.1 | eV |
| Electrical field | $E$ | $-5 \times 10^4$ | $V.m^{-1}$ |

| Temperature gradient | $\kappa$ | $5 \times 10^9$ | $K \cdot m^{-1}$ |

In Fig.1, we plot the charge current density as a function of position x and y. The variation of charge current with respect to position and time is governed by the continuity equation of charge density and charge current density (12). For simplicity, we only study the charge current at steady state. We can see that the charge current density decreases gradually along both the x and y directions, which is due to the resistance in the ferromagnet, in our calculation it is concerned with the momentum relaxation time $\tau$ of electrons. Since the external electric field is applied only along the x-direction, the variation of charge current along y-direction is mainly caused by the Dresselhaus SOC.

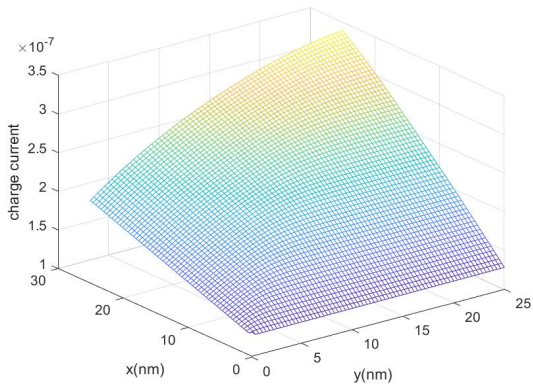
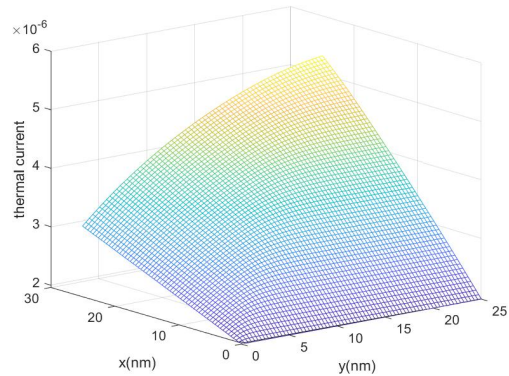

Fig.1 The charge current density vs position

Fig.2 The thermal current density vs position

We also show the curve of thermal current density as a function of position in Fig. 2, which is similar to the charge current density because of their definitions, it also decreases with position gradually. Here we only consider the thermal current density carried by the transport electrons. Besides the external electric field, the thermal current density can also be driven by the temperature gradient. Since the electric field and gradient of temperature are all along the x-axis, the variation of thermal current density along y-direction is primarily induced by the Dresselhaus SOC.

Because we choose the magnetization of ferromagnet $\vec{M} = (0,0,1)$, so the z component of STT is 0. In Fig.3, we show the x and y components of STT density as a function of position. The usual STT is the space integral of this density over the rectangular ferromagnet. It is shown that the magnitude of STT density is different at different position for both the x and y components. The magnetization of ferromagnet at $x = 10nm$ is easiest to be switched by the bigger STT, and is hardest to flip around the line $y = x$ because of the smaller STT. The switching of magnetization will produce the spin wave within the two- dimensional ferromagnet.

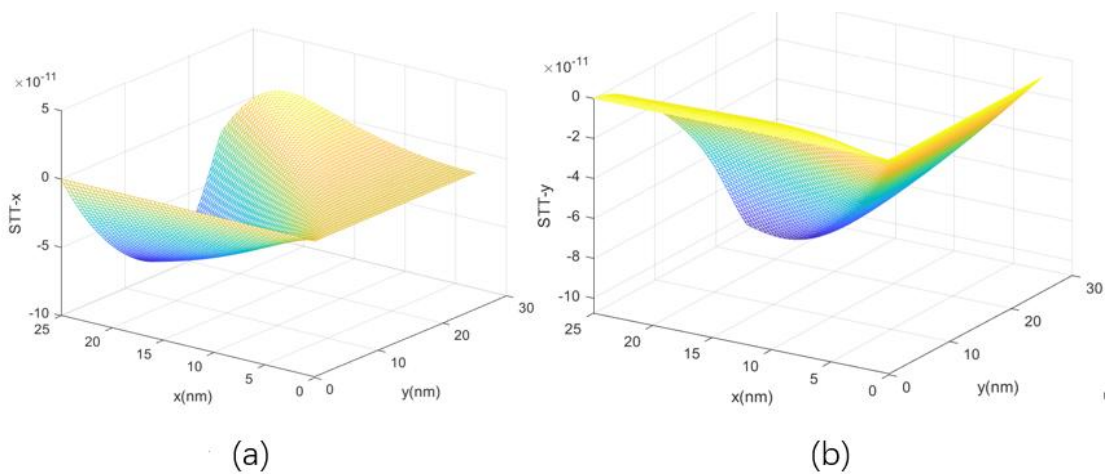

(a)          (b)

Fig.3 (a) The x-component of STT density. (b) The y-component of STT density

The spin current density vs position is shown in Fig.4. Since the spin current is a tensor, we only draw the xx-, xy- and xz-components of spin current density as a function of position, they vary obviously around the line $y = x$. The variation of spin current with position and time satisfies the continuity equation (13) for the spin accumulation and spin current. According to Eq. (14), the divergence of spin current will make a contribution to the Zhang-like STT [28].

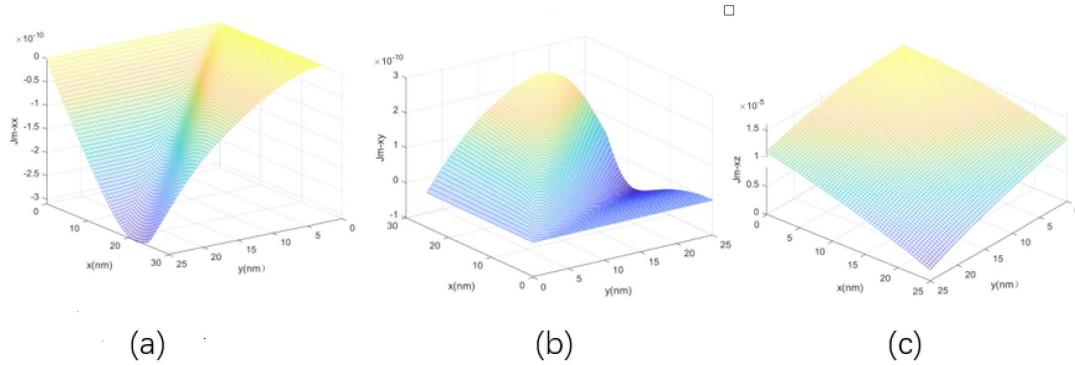

Fig. 4 (a) The xx- component of spin current (b) The xy- component of spin current (c) The xz- component of spin current

In Fig.5, we draw the SOT as a function of position. The usual SOT is expressed as $\frac{\lambda_D}{\hbar}(\vec{e}_x \frac{\partial \rho}{\partial x} - \vec{e}_y \frac{\partial \rho}{\partial y}) - \frac{\lambda_D}{\hbar^2} \int (p_x \vec{e}_x - p_y \vec{e}_y) \times \vec{g}\, d\vec{p}$, so it is not sensitive to the temperature. It decreases along x-direction obviously, while there is small variation along y-direction, because the external electric field is applied along x-direction. For comparison, we also plot the TSOT at different temperature 300K, 200K and 100K in Fig.6, respectively, which depend on the temperature and it's gradient obviously. The higher of temperature, the bigger of TSOT, because there are more polarized electron participating in transport at higher temperature. Compared Fig.6 with Fig.5, we can find that the TSOT is smaller than the usual SOT, while the TSOT can become bigger when we increase the temperature, so the TSOT can not be negligible at higher temperature. Certainly, TSOT is also proportional to the temperature gradient, it will play an important role at higher temperature gradient. It should be pointed out that the TSOT is calculated by Eq. (15), we only need the expression of local equilibrium distribution function, it is very simple than Freimuth's expression of Berry phase [20-21], because the latter need the electronic wavefunction obtained usually by DFT. By means of Eq. (15), we can calculate the TSOT easier, this is the advantage of SBE method.

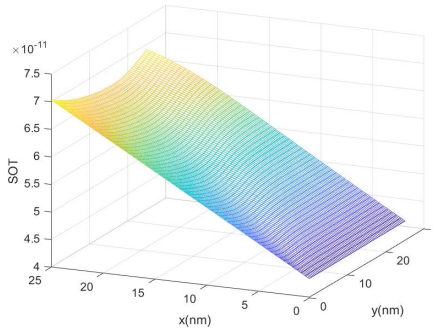 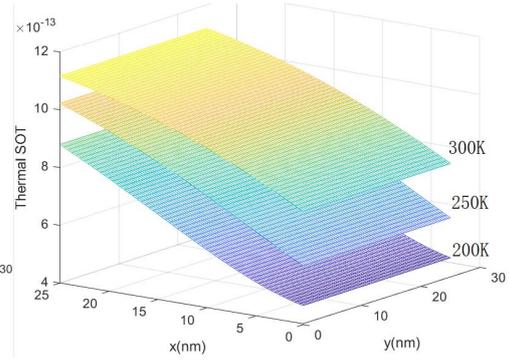

Fig. 5 The SOT vs position.   Fig.6 The TSOT vs position at different temperature T=300K, 200K, 100K.

## IV. Summary and Discussions

In this paper, we have derived the TSOT in a two-dimensional ferromagnet with Dresselhaus SOC by SBE under the local equilibrium assumption. The usual SOT is induced by the external electric field applied to the system, while the TSOT is driven by the gradient of temperature. We also find that TSOT is very sensitive to the temperature, the higher temperature, the bigger TSOT. Our results show that the TSOT is smaller than SOT, but it can not be negligible at higher temperature. Certainly, TSOT is also proportional to the temperature gradient, according to its expression Eq. (15). Because the direct experiment to observe TSOT in two-dimensional ferromagnets with Dresselhaus spin-orbit coupling haven't been carried out now, we only predict theoretically that one can observe the effects of TSOT in the case of big gradient of temperature and higher temperature.

To simplify our calculation, we only choose a simple uniform magnetization in a 2-dimensional ferromagnet, while in reality the magnetization usually varies with time and position. The variation of magnetization would have influence on the transport properties of the spin-polarized electrons. If we consider the variation of magnetization, the calculation will become much more complicated, it is left for future exploration.

## Acknowledgments

This study is supported by the National Key R&D Program of China (Grant No. 2022YFA1402703), the Strategic Priority Research Program of the Chinese Academy of Sciences (Grant No. XDB28000000). We also thank Prof. Gang Su, Zhen-Gang. Zhu, Bo.

Gu and Qing-Bo. Yan for their helpful discussions.

**Data Availability Statement**

Data sets generated during the current study are available from the corresponding author on reasonable request.